\documentclass[11pt]{article}
\usepackage{psfig}
\usepackage{amsmath}
\usepackage{geometry}                
\usepackage{graphicx}
\usepackage{amssymb}
\usepackage{epstopdf}

\usepackage {natbib}

\usepackage{epsfig}
\usepackage{url}

\begin{document}

\fontsize{11}{14.5pt}\selectfont


\begin{center}

\vspace*{0.9in}

{\Large\bf Bayesian Gene Set Analysis}\\[16pt]

 \begin{center}
{\large Babak Shahbaba \footnote[1]{Department of Statistics, University of California at Irvine, Irvine, CA, USA, \texttt{babaks@uci.edu}}, Robert Tibshirani \footnote[2]{Department of Statistics, Stanford University, Stanford, CA, USA, \texttt{tibs@stanford.edu}}, Catherine Shachaf  \footnote[3]{Department of Microbiology and Immunology, Stanford University, Stanford, CA, USA, \texttt{cshachaf@stanford.edu}}, and Sylvia K. Plevritis \footnote[4]{Department of Radiology, Stanford University, Stanford, CA, USA, \texttt{sylvia.plevritis@stanford.edu}}}\\[2pt]

 \end{center}

\date{}

\end{center}

\vspace*{8pt}

\noindent \textbf{Abstract.}
{Gene expression microarray technologies provide the simultaneous measurements of a large number of genes. Typical analyses of such data focus on the individual genes, but recent work has demonstrated that evaluating changes in expression across predefined sets of genes often increases statistical power and produces more robust results. We introduce a new methodology for identifying gene sets that are differentially expressed under varying experimental conditions. Our approach uses a hierarchical Bayesian framework where a hyperparameter measures the significance of each gene set. Using simulated data, we compare our proposed method to alternative approaches, such as Gene Set Enrichment Analysis (GSEA) and Gene Set Analysis (GSA). Our approach provides the best overall performance. We also discuss the application of our method to experimental data based on p53 mutation status.}

\section{Introduction}
High throughput biological experiments such as gene expression microarray analysis have traditionally focused on identifying individual genes that are differentially expressed across experimental conditions \citep[see for example,][]{baldi01, efron01, newton01, tusher01, gottardo03, ishwaran03, ishwaran05, smyth04, delmar05a, do05, opgen07}. These studies typically involve simultaneous analysis of tens of thousands of candidate genes with relatively small number of samples, often much less than 100. While statistical methods developed for analyzing such data are easy to implement, interpretation of the resulting list of significant genes is far from trivial. As stated by \cite{subramanian05}, it is not uncommon for such studies to either produce a large list of significant candidates with no unifying biological theme or identify no significant gene at all (e.g., after correcting for multiple hypothesis testing in a frequentist framework). Moreover, on many occasions, there is little overlap between findings of different research groups for the same biological system \citep{subramanian05}; that is, the results have low reproducibility. These issues are partly due to the oversimplification of the problem (e.g., ignoring the interconnectivity of genes) in order to make statistical analysis of data feasible. 

In recent years, several methods have been proposed to overcome these limitations by incorporating biological knowledge in the data analysis. These methods produce inferences on predefined sets of genes as opposed to individual genes \citep{virtaneva01, pavlidis02, mootha03, smyth04, rahnenfuhrer04, subramanian05, barry05, newton07, efron07}. The gene sets are defined based on prior biological information such as biochemical pathways or similarity of gene sequences. Shifting the focus of analysis from individual genes towards groups of genes often increases statistical power and leads to more reproducible results \citep{efron07}. 

Typically, gene set analysis methods start with a collection of gene sets $\mathcal{S}_1$, $\mathcal{S}_2$, \ldots, $\mathcal{S}_K$ defined based on some prior information. Then, for each individual gene, a statistic, $z$, (e.g., two sample t-statistic), is calculated to measure the amount of support for its corresponding hypothesis. Then, for each gene set $\mathcal{S}_{s}$, a set-level statistic $T_{s}$ is calculated as a function of individual statistics ${\bf{z_{s}}} = \{z_{s1}, z_{s2}, ..., z_{s \ell_{s}}\}$, where $\ell_{s}$ is the number of genes assigned to that set. The significance of $T_{s}$ is assessed by using a permutation test where the class labels are repeatedly permuted. 

One of the most widely used methods for analyzing gene sets is Gene Set Enrichment Analysis (GSEA). The gene-level test statistic, $z$, in GSEA is calculated using $t$-test when the analysis involves two classes (e.g., treatment vs. control). The set-level statistic for each set is defined as a signed Kolmogorov-Smirnov (KS) statistic based on the distribution of ${\bf{z}_{s}}$ and that of its complement, ${\bf{z}_{s}^{c}}$, which includes all genes that are not included in the set $s$. Recall that the KS statistic quantifies the distance between the empirical cumulative distribution function of a sample (in this case, the gene set of interest) and that of a reference sample (in this case, the complement set). The sign represents the direction of shift in the distribution. 

The GSEA method poses several statistical concerns \citep{damian04, efron07}. For example, \cite{efron07} argue that while GSEAÕs dependence on KS is reasonable, it is not necessarily the best choice. This is because KS is sensitive to change in the distribution of $z$ scores across the whole distribution, whereas we are interested mainly in changes in lower and upper distribution tails, where genes are statistically significant. \cite{efron07} propose an alternative approach that has a significant power advantage over GSEA. Their method uses the a new statistic called \emph{maxmean}. For each gene set, maxmean is calculated by first separating negative and positive $z$-scores (i.e., separating downregulated and upregulated genes within the set). Then, we set the positive $z$-scores to 0 and take the average over all scores (zeros included). We denote this average as $T^{-}$. Next, we set all negative $z$-scores to 0 and take the average over all scores (zeros include). We denote this average as $T^{+}$. The maxmean statistic, $T_{max}$, is the larger average in absolute value; that is, $T_{max} = \max(|T^{-1}|, |T^{+}|)$.

In their paper, \cite{efron07} also argue that any method for assessing gene sets should compare a given gene set score not only to scores from permutations of the sample labels, but also to scores from sets formed by random selections of genes. To illustrate this concept, they generated data on 1000 genes and 50 samples, with each consecutive non-overlapping block of 20 genes considered to be a gene set. The first 25 samples are the control group, and the second 25 samples are the treatment group. First they generated each data value independently from $N(0, 1)$, and then they increased the expression values for the treatment group by the constant 2.5 for the first 10 genes in each gene set. This way, all of the maxmean scores look significantly large compared to the permutation values. But they argue that given the way that the data were generated, there seems to be nothing special about any one gene set. To address this issue, they propose \emph{restandardization}, where they center and scale the maxmean statistic by its mean and standard deviation under the randomization of genes among gene sets. This way, no gene set would be selected as significant. 

Although the maxmean statistic has several advantages (e.g., higher statistical power) over other statistics, it has some limitations. More specifically, we believe that the maxmean statistic performs well when almost all significant candidate genes in a set are either upregulated or downregulated. To illustrate this, consider two gene sets, $\mathcal{S}_1$ and $\mathcal{S}_2$, each with 100 genes. In the first set, 50 genes have $z$-scores equal to $-2$, and the $z$-scores of the remaining genes are zero. The second gene set also has 50 $z$-scores of $-2$; however, the remaining genes have $z$-scores equal to $1.5$. While it is possible that for a given experiment one might regard $\mathcal{S}_2$ as a more significant gene set compared to $\mathcal{S}_1$, the maxmean score for both gene sets is 1. In general, maxmean mainly captures changes in the dominating regulation direction, ignoring (to some extent) when differentially expressed genes within a set do not necessarily move in the same direction.    
 
In this paper, we propose an alternative hierarchical Bayesian method for gene set analysis. In the next section, we describe our method in detail. In Section 3, we use simulated data to evaluate the performance of our model. We compare our model to several other approaches including GSEA \citep{subramanian05} and GSA \citep{efron07}. Section 4 shows the results from applying our model to real data. Finally, Section 5 is devoted to discussion and future directions.
  
\section{Bayesian analysis of gene sets}
We denote the $i^{th}$ observed expression of gene $g$ in set $s$ as $y_{sgi}$. Given the class label $x_i$, where $x_i \in \{0, 1\}$ (i.e., control=0, treatment=1), we assume the following model for $y_{sgi}$:
\begin{eqnarray*}\label{model1}
y_{sgi} & = & \alpha_{sg}+\beta_{sg} x_i + \epsilon_{sgi} \qquad g=1, 2, \ldots, \ell_{s} \\
\epsilon_{sgi} & \sim & N(0, \sigma^{2}_{sg})
\end{eqnarray*}
Here, $\alpha_{sg}$ is the overall mean of expression for gene $\mathcal{G}_{g}$ in the set $\mathcal{S}_{s}$, and $\beta_{sg}$ is the expected change in the expression of this gene between the control group (i.e., $x_{i} = 0$) and the treatment group (i.e., $x_{i} = 1$). The model therefore can be considered as a simple linear mixed-effects model, where $\beta_{sg}$ are the random effects associated with the class variable. We could therefore use a mixed effects ANOVA model to evaluate the overall significance of a gene set, where the null hypothesis is defined as $H_{0}: \beta_{s1}=\ldots=\beta_{s\ell_{s}}=0$. Alternatively, assuming that $\beta_{sg} \sim N(0, \tau^{2}_{s})$, we have $H_{0}: \tau^{2}_{s} = 0$. Standard computer packages such as \emph{nlme} and \emph{lme4} in R could be used for such analysis. Note however that the clustering of observations is based on genes (i.e., variables) as opposed to subjects (which is more common in mixed-effects models). 

\subsection{A hierarchical Bayesian model}\label{BGSA}
A simple mixed-effects model with a fixed intercept and a random effect parameter is equivalent to a simple hierarchical Bayesian model with uniform prior on the intercept and a normal prior (with mean 0 and unknown variance) on the random effect parameter. However, compared to mixed-effects models, hierarchical Bayesian models are more flexible, and as discussed in this paper, they can easily handle more complex data structure such as overlaps between gene sets. Using appropriately informative (i.e., reasonably broad) hyperpriors, these models can also accommodate shrinkage of $\beta$'s and their corresponding variance $\tau^{2}_{s}$ towards zero when the gene set as a whole is not significant. This way, they avoid strong influences of one or two differentially expressed genes on the overall measure of significance for the set. Moreover, by using more flexible priors, such as the mixture prior proposed in the in this paper, we can mitigate the negative effect of wrong model assumptions (e.g., normality of random effects when they are not normally distributed). Therefore, we believe that models with random effects are more naturally discussed within the hierarchical Bayesian framework where by default all parameters are considered to be random (in general, there is no fixed effect). Here, we use the following prior distributions for the parameters of our model:
\begin{eqnarray*}
\sigma^{2}_{sg} | \xi, \eta & \sim & \textrm{Inv-}\chi^{2}(\xi, \eta^{2})\\
\alpha_{sg} | \gamma & \sim & N(0, \gamma^{2}) \\
\beta_{sg} | \tau & \sim & N(0, \tau^{2}_{s})
\end{eqnarray*}
Note that the priors for $\sigma^{2}_{sg}$ and $\alpha_{sg}$ are the same for all genes regardless of their associated gene sets. We could set $\xi, \eta$ and $\gamma$ to some constant values such that the resulting distributions are appropriately broad giving reasonable prior probability to the set of possible values. For simplicity, however, we use non-informative priors for this parameters. To this end, we assume $P(\alpha_{sg}, \sigma^{2}_{sg}) \propto 1/ \sigma^{2}_{sg}$. For $\sigma^{2}_{sg}$, this is equivalent to setting $\xi=0$ and $\eta=1$. For $\alpha_{sg}$, this is equivalent to the limit of the normal distribution when we take $\gamma^{2} \to \infty$.

For the prior of $\beta_{sg}$, we regard $\tau^{2}_{s}$ as hyperparameter with its own hyperprior,  $\tau^{2}_{s} \sim \textrm{Inv-}\chi^{2}(\nu, \phi^{2})$. Note that $\tau^2_{s}$ is a shared by all genes in the set $\mathcal{S}_{s}$ (throughout this paper, we assume that gene expression values are normalized). The role of $\tau^2_{s}$ is to adjust the distribution of $\beta$'s within a gene set. If a gene is not differentially expressed, its corresponding $\beta$ moves close to zero in posterior. This will be accommodated by small values of $\tau^{2}_{s}$. If a large number of genes in a set are not differentially expressed (i.e., negligible amounts of shift are observed), the hyperparameter $\tau^2_{s}$, which is shared by all the genes in the set, would shrink towards zero.  If a gene is in fact significant, we expect a large shift in its expression value for the case group, which in turn makes the posterior distribution of $\beta$ move away from zero. When the gene set includes a large number of significant genes, $\tau^{2}_{s}$ becomes larger in posterior to allow for large values of $\beta$. Therefore, we can evaluate the overall significance of a gene set based on the posterior distribution of $\tau^{2}_{s}$: for a less significant gene set, the posterior distribution of $\tau^{2}$ shrinks towards zero, whereas for a significant gene set, the posterior distribution would higher probabilities to large values of $\tau^{2}$. Similarly, we can evaluate significance of individual genes within a set using the posterior distribution of their corresponding $\beta$: the posterior distribution of $\beta$ for a more significant gene would be away from zero (we can quantify significance of gene by measuring the posterior tail probability of zero).

\begin{figure}
\begin{center}
 \centerline{\includegraphics[width=5in]{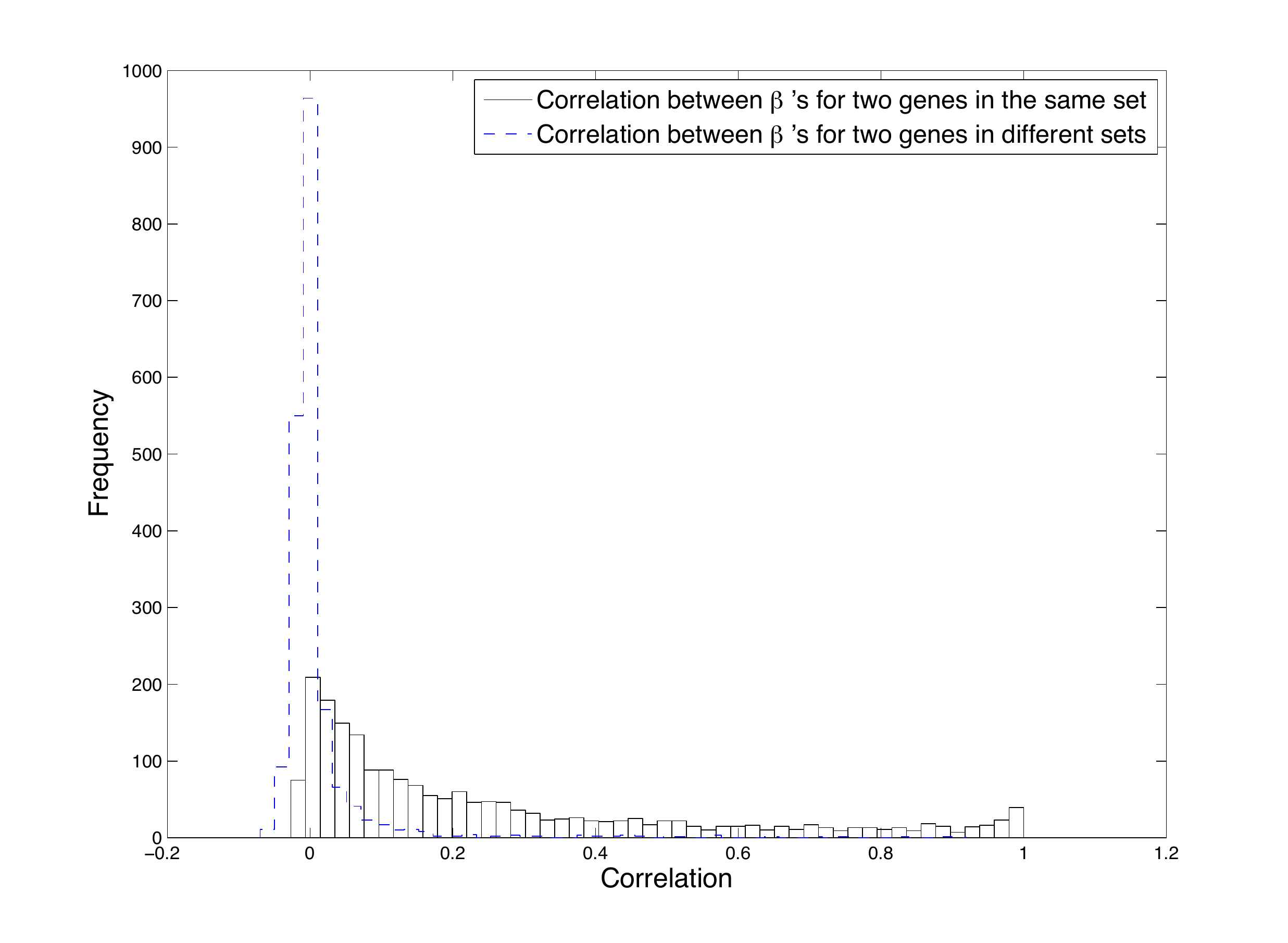}} 
   \vspace*{-12pt}
\caption{Comparing the histogram of correlation between absolute values of $\beta$'s for two genes in one set (solid line) to the corresponding histogram of the correlation between the absolute values of $\beta$'s for genes in two different sets (dashed line).} %
 \label{corrResOneBeta} %
\end{center} 
\end{figure}

This hierarchical Bayesian model introduces prior correlation between the absolute values of $\beta$'s for genes within one set, while $\beta$'s associated with genes in different sets remain independent. We illustrate this by showing samples simulated from the above hierarchical prior. Consider two gene sets each with two genes whose expressions are measured under two different experimental conditions. We generate samples from the prior as follows. For each gene set, we first sample $\eta \sim \textrm{Inv}-\chi^{2}(1, 0.5)$. Next, given the current value of $\eta$, we sample $\tau_1$ and $\tau_2$ from $\textrm{Inv}-\chi^2(1, eta)$. Finally, we sample 100 $\beta$'s from $N(0, \tau)$ and measure the correlation between the absolute values of $\beta$'s for genes in one set (denoted as $r_1$) as well as the correlation between the absolute values of $\beta$'s for genes in two different gene sets (denoted as $r_0$). We repeat this procedure 1000 times. The histograms for $r_0$ and $r_1$ are presented in Figure \ref{corrResOneBeta}. As shown, the correlation between $|\beta|$'s tends to be higher for genes within the same set compared to that of genes in different sets. Therefore, if the gene sets are defined properly, we expect that the informative prior incorporated in our model results in better performance in terms of identifying significant pathways.   

To illustrate the above approach, we use a simple simulation study. We first generate a dataset with 1000 candidate genes and 30 samples such that the first 15 samples are assigned to the control group, and the rest to treatment group. We assume that each consecutive non-overlapping block of 20 genes forms a set; that is, there are 50 gene sets in total. The values for all genes are first sampled independently from $N(0, 1)$. To make a gene significant, we add then the constant 1 to the expression values for the treatment group. In Set 1 we make all 20 genes significant. In Set 2, we make only half of the genes (i.e., 10) significant. We reduce the number of significant genes to 5 for Set 3 and to 2 for Set 4. We keep the remaining 46 gene sets (i.e., Sets 5-50) as non significant. Therefor, the significance of gene sets reduces from Set 1 to Set 4, and there is no association between the class variable and Sets 5-50. We use our hierarchical Bayesian method to evaluate the significance of these gene sets.

\begin{figure}
\begin{center}
 \centerline{\includegraphics[width=3in]{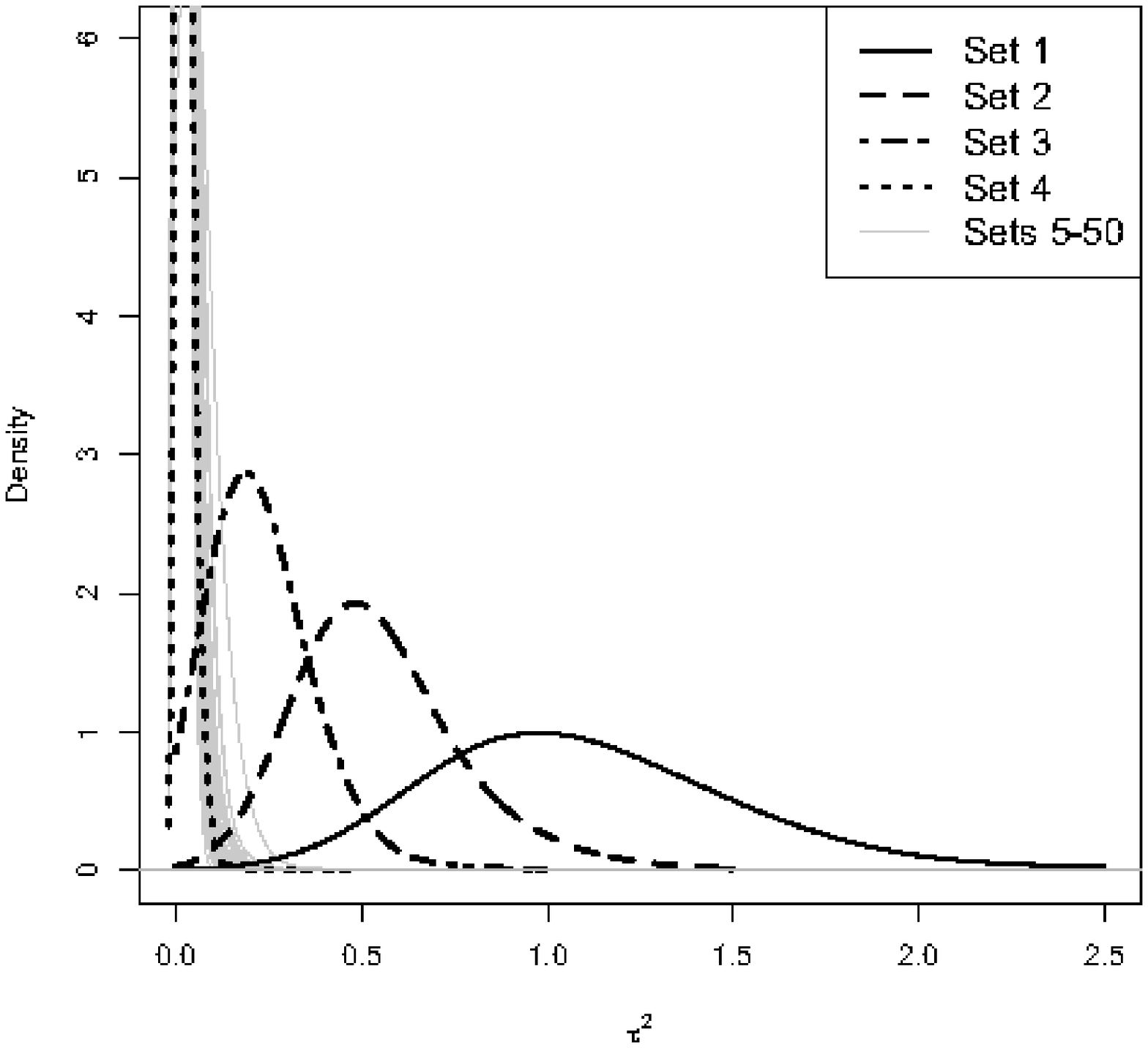} \includegraphics[width=3in]{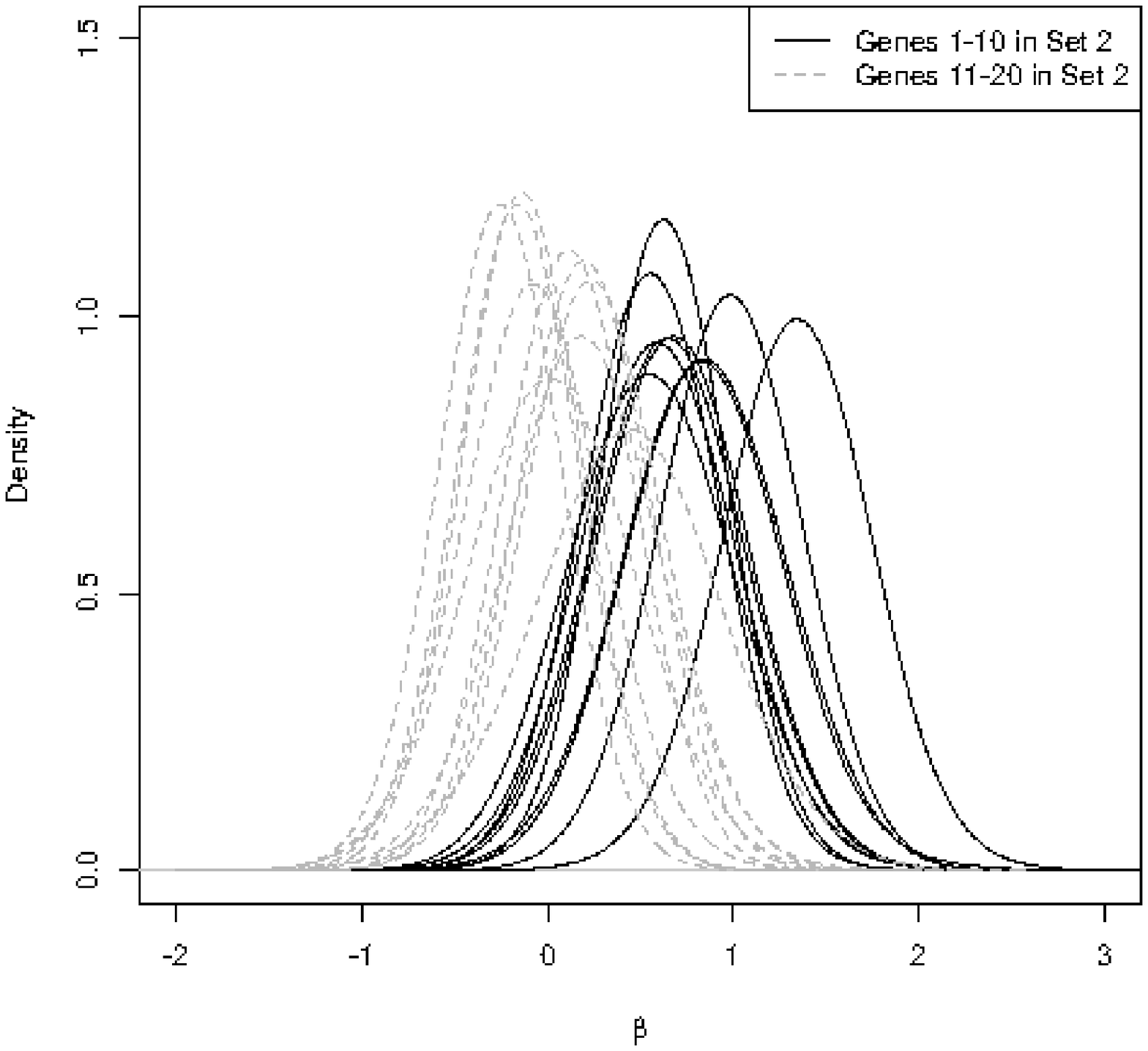}} 
   \vspace*{-12pt}
\caption{Left panel: Posterior distribution of the hyperparameter $\tau^{2}$ for 50 simulated gene sets. Sets 1-4 have 20, 10, 5, and 2 significant genes respectively. Sets 5-50 contain no significant genes. Right panel: Posterior distribution of $\beta$ for individual genes in Set 2, where only 10 genes are significantly upregulated (the constant 1 is added to the expression values for the treatment group).}
\label{postTau2}
\end{center}
\end{figure}

Figure \ref{postTau2} shows the posterior distribution of $\tau^{2}$ for the 50 gene sets. As we can see, there is a clear separation between the posterior distribution of $\tau^{2}$ for Set 1 and the posterior distributions of $\tau^{2}$ for Sets 5-50. As the number of significant genes reduces from Set 1 to Set 4, the posterior distribution of $\tau^{2}$ shrinks towards 0. The posterior distribution of $\tau^{2}$ for Set 4 (where only 2 genes out of 20 are significant) is very similar to the posterior distribution of $\tau^{2}$ for Sets 5-50.  The posterior expectation of $\tau^2$ for Sets 1 to 4 are 1.11, 0.55, 0.23, and 0.02 respectively. The posterior expectation of $\tau^2$ for Sets 5-50 ranges from 0.02 to 0.05.

As mentioned above, we can use the posterior distribution of $\beta$'s to evaluate the significance of individual genes. Figure \ref{postBeta} shows the posterior distribution of $\beta$ for individual genes in Set 2 (where only 10 genes out of 20 are significant). As we can see, for the significant genes, the posterior distributions of $\beta$ moves away from 0 (in this case to the right since genes are upregulated). 

\subsection{Gene set selection via mixture priors}\label{BGSAmix}
The above approach could be useful for ranking gene sets in terms of their significance based on, for example, posterior expectations of $\tau^{2}$; that is, the higher the posterior expectation of $\tau^{2}$, the more significant the gene set. However, in many practical problems, an automatic mechanism for selecting significant gene sets (or genes) is required. For linear mixed-effects models, we could obviously use $p$-values with a cutoff that has been adjusted to control the family-wise error rate (i.e., the chance of making at least one type I error) or false discovery rate (i.e., the expected proportion of false positives among the rejected null hypotheses). For our model, we do this by using a mixture prior, similar to that of \cite{george93} and \cite{ishwaran05}, instead of a simple scaled-inv-$\chi^2$ distribution as the prior of $\tau^{2}_{s}$. That is, we assume
\begin{eqnarray*}
\tau^{2}_{s} & \sim & (1-\lambda)F_{0} + \lambda F_{1},
\end{eqnarray*}
where $F_{0}$ and $F_{1}$ are the distributions of $\tau^{2}_{s}$ under the null and alternative hypotheses respectively, and $\lambda$ is the probability that $\tau^{2}_{s}$ is generated under the alternative. As mentioned before, under the null hypothesis of no relationship between a gene set and the class variable, we have $\tau^{2}_{s} = 0$. Therefore, we could assume that under the null $F_0 \equiv  \delta_{0}$ (i.e., point mass distribution at 0). However, making such assumption is appropriate if we are using the correct model (e.g., all the relevant factors are included in the model), and data in fact conform with the model assumptions. This rarely happens in reality. A more reasonable assumption is that under the null $\tau^{2}_{s}$ are quite small and close to zero. Therefore, we assume the following mixture prior for $\tau^{2}_{s}$:
\begin{eqnarray*}
\tau^{2}_{s}|\lambda, \phi_0, \phi_1 & \sim & (1-\lambda)  \textrm{ Inv-}\chi^2(\nu, \phi^{2}_{0}) +  \lambda  \textrm{ Inv-}\chi^2(\nu, \phi^{2}_{0} + \phi^{2}_{1})\\
\phi^{2}_{0}, \phi^{2}_{1} & \sim & \textrm{Gamma}(1, 1)\\
\nu & \sim &  \textrm{Gamma}(1, 1) \\
\lambda & \sim & \textrm{Beta}(1, 1)
\end{eqnarray*}
This way, the distribution of $\tau^{2}_{s}$ are assumed to come from two different distributions: a scaled-inv-$\chi^{2}$ distribution with a relatively small scale parameter $\phi^{2}_{0}$, and a scaled-inv-$\chi^{2}$ with a relatively larger scale parameter $\phi^{2}_{0}+\phi^{2}_{1}$. As shown in Figure \ref{ichi2}, for a given degrees of freedom, as the scale parameter increases, scaled-inv-$\chi^{2}$ distribution gives higher probability to large values of random variable (i.e., the distribution moves to the right). Therefore, our prior separates $\tau^{2}_{s}$ into two groups. One group includes small values of $\tau^{2}$ with Inv-$\chi^2(\nu, \phi^{2}_{0})$ distribution. These $\tau^{2}_{s}$ are assumed to be generated under the null hypothesis. The second group includes large values of $\tau^{2}$ with  Inv-$\chi^2(\nu, \phi^{2}_{0}+\phi^{2}_{1})$. These $\tau^{2}_{s}$ are assumed to be generated under the alternative hypothesis. 

\begin{figure}
\begin{center}
 \centerline{\includegraphics[width=3in]{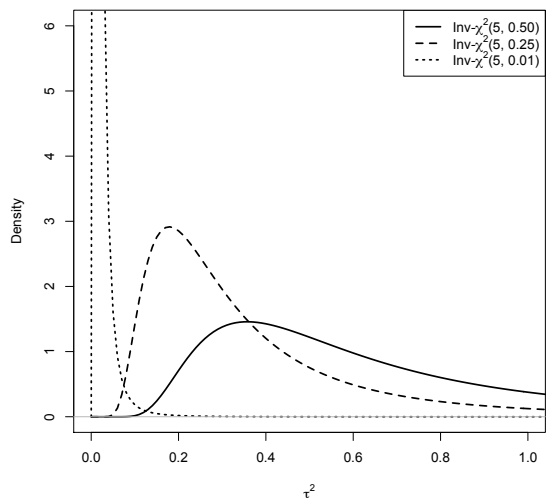} }
 \vspace*{-12pt}
\caption{Three scaled-inv-$\chi^2$ distributions with the same degrees of freedom and different scale parameters. As the scale parameter increases, the distribution moves to the right giving higher probabilities to large values of $\tau^2$.} 
\label{ichi2}
\end{center}
\end{figure}

For $\lambda$, we use a conjugate Beta prior. Throughout this paper, we assume $a=b=1$, which is equivalent to the Uniform(0, 1) distribution, which is a noninformative prior. In practice, we might expect that only a small number of gene sets are significant and use informative priors such as Beta$(1, 10)$. 

To facilitate posterior sampling, we use the data augmentation approach \citep{tanner87} by introducing binary latent variables $v_s \sim \textrm{Bernoulli}(\lambda)$ such that
\begin{eqnarray*}
\tau^{2}_{s} |v_s, \phi_0, \phi_1 & \sim & (1-v_s) \textrm{ Inv-}\chi^2(\nu, \phi^{2}_{0}) +  v_s  \textrm{ Inv-}\chi^2(\nu, \phi^{2}_{0} + \phi^{2}_{1}).
\end{eqnarray*}
Alternatively, we can write the above prior as
\begin{eqnarray*}
\left\{ \begin{array}{l} 
\tau^{2}_{s}|v_{s}=0, \nu, \phi_0 \sim \textrm{Inv-}\chi^{2}(\nu, \phi^{2}_{0})\\ 
\tau^{2}_{s}|v_{s}=1, \nu, \phi_0, \phi_1 \sim \textrm{Inv-}\chi^{2}(\nu, \phi^{2}_{0} + \phi^{2}_{1})
\end{array} \right.
\end{eqnarray*} 
Note that we can use the posterior distribution of $v_s$ given the observed data, $D$, to evaluate the significance of each gene set. To this end, we use $P(v_s = 0|D)$ (integrated over all other parameters) as an estimate for the posterior probability of $H_0$ given the data, $P(H_0|D)$. 
 
We applied this model to the simulated data discussed above. For Sets 1-4, the estimated values for $P(v_s = 0|D)$ are 0.002, 0.008, 0.06, and 0.99 respectively. For Sets 5-50 (non significant genes sets) this probability ranges from 0.96 to 1. 
\begin{figure}
\begin{center}
 \centerline{\includegraphics[width=2.5in]{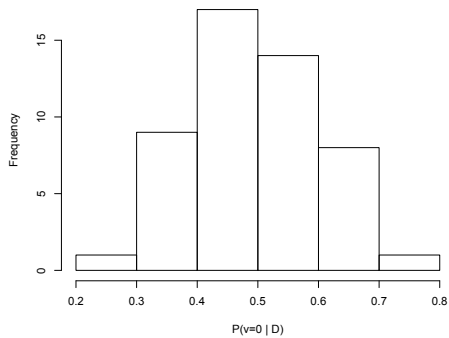} }
  \vspace*{-12pt}
\caption{Histogram of $P(v = 0 |D)$ for simulated data where all gene sets are equally upregulated such that no gene set is special.} 
\label{ichi2}
\end{center}
\end{figure}

One advantage of our model is that it automatically performs a form of what \cite{efron07} called restandardization. To show this, we generate data similar to the illustrative example above, but this time we increase the expression values for the treatment group by 1 unit for half of the genes in every gene set. This way, we make half of the genes in each set upregulated, but there is nothing special about any of the gene sets. Figure \ref{histAllSignificant} shows the histogram of  $P(v = 0|D)$ for this data set. Notice that the values of $P(v = 0|D)$ are centered around 0.5. They range from 0.29 to 0.77, which is consistent with the fact that no set is special. 
\subsection{Posterior sampling}\label{posterior}
The priors used in this paper are all conditionally conjugate except for the priors used for scale parameters of scaled-inv-$\chi^{2}$ distributions (e.g., $\nu$, $\phi^{2}_{0}$ and $\phi^{2}_{1}$). Using conjugate priors make posterior sampling easier since posterior distributions are conditionally tractable and we can use the Gibbs sampler. The posterior distributions for $\sigma^{2}_{sg}$ and $\alpha_{sg}$ given all other parameters are as follows:
\begin{eqnarray*}
\qquad \sigma^{2}_{sg} | . & \sim & \textrm{Inv-}\chi^{2} \Big(n, \frac{\sum_{i=1}^{n}(y_{sgi} - \alpha_{sg} - \beta_{sg}x_{i})^{2}}{n} \Big),\\
\alpha_{sg} | . & \sim & N\Big(\frac{   \sum_{i}^{n} (y_{sgi} - \beta_{sg} x_{i}) }{n}, \frac{\sigma^{2}_{sg}}{n} \Big).
\end{eqnarray*}
Samples from the posterior distribution of $\beta_{sg}$ obtained given the updated values of the above parameters as well as the current value of $\tau^{2}_{s}$,
\begin{eqnarray*}
\beta_{sg} |. & \sim & N \Big(\frac{  \frac{1} { \sigma^{2}_{sg}}  \sum_{i}^{n} (y_{sgi}  - \alpha_{sg} ) x_i } {\frac{1}{\tau^{2}_{s}} + \frac{ \sum_{i}^{n} x_i}{\sigma^{2}_{sg}} }, \Big[\frac{1}{\tau^{2}_{s}} + \frac{ \sum_{i}^{n} x_{i} } { \sigma^{2}_{sg} }\Big]^{-1} \Big)
\end{eqnarray*}
The posterior distribution of $\tau^{2}_{s}$ has a closed form given the current values of $\beta$'s, $v_{s}$, $\nu, \phi^{2}_{0}$ and $\phi^{2}_{1}$, 
\begin{eqnarray*}
\left\{ \begin{array}{l} 
\tau^{2}_{s}|v_{s}=0, \nu, \phi^{2}_{0}, \beta  \sim  \textrm{Inv-} \chi^{2} \Big(\nu+\ell_{s},  \frac{ \nu \phi^{2}_{0} + \sum_{g=1}^{\ell_{s}} (\beta_{sg})^{2} } {\nu+\ell_{s}} \Big), \\ 

\vspace*{6pt}

\tau^{2}_{s}|v_{s}=1, \nu, \phi^{2}_{0},  \phi^{2}_{1}, \beta  \sim  \textrm{Inv-}\chi^{2}\Big (\nu+\ell_{s},  \frac{\nu(\phi^{2}_{0}+\phi^{2}_{1}) + \sum_{g=1}^{\ell_{s}}(\beta_{sg})^{2} }{\nu_{0}+\ell_{s}} \Big).
\end{array} \right.
\end{eqnarray*} 
We sample $v_{s}$ from a Bernoulli distribution with the following probability:
\begin{eqnarray*}
P(v_{s} = 1 | \tau^{2}_{s},  \nu, \phi^{2}_{0}, \phi^{2}_{1}, \lambda)  =  \frac{\lambda f(\tau^{2}_{s}|\nu, \phi^{2}_{0}+\phi^{2}_{1}) } {(1-\lambda) f(\tau^{2}_{s}|\nu, \phi^{2}_{0})+ \lambda f(\tau^{2}_{s}|\nu, \phi^{2}_{0}+\phi^{2}_{1})}, 
\end{eqnarray*}
where $f(\tau^{2}_{s}|\nu, \phi^{2}_{0})$ and $f(\tau^{2}_{s}|\nu, \phi^{2}_{0}+\phi^{2}_{1})$ are densities for scaled-inv-$\chi^{2}$ distributions. The conditional posterior distribution of $\lambda$ also has a closed form, 
\begin{eqnarray*}
\lambda & \sim & \textrm{Beta}(a+\sum_{s}^{K}v_{s}, b + K - \sum_{s}^{K}v_{s}).
\end{eqnarray*}
Here, $K$ is the total number of gene sets. 

Because priors for $\nu, \phi^{2}_{0}$ and $\phi^{2}_{1}$ are not conditionally conjugate, we cannot use the Gibbs sampler for these parameters. Note that the computational burden is negligible. To sample from the posterior distribution of $\nu, \phi^{2}_{0}$ and $\phi^{2}_{1}$, we use single-variable slice sampling \citep{neal03}. At each iteration, we use the "stepping out" procedure to find the interval around the current point and the "shrinkage" procedure for sampling from the interval. 

\section{Results for simulated datasets}
We conduct six simulation studies in order to evaluate the performance of our approach, henceforth called Bayesian Gene Set Analysis (BGSA), in identifying significant sets. More specifically, we compare our BGSA model to a linear mixed-effects (LME) model, GSEA \citep{subramanian05}, maxmean statistic \citep{efron07}, and two other summary statistics based on the mean of $z_i$ and the mean of $|z_i|$, which we refer to as Mean.z and Mean.abs.z respectively. For the LME model, we use \emph{lme4} package in R. We evaluate the significance of each gene set by comparing two nested models. The first model is a full model that includes the intercepts $\alpha_{sg}$ and random coefficients $\beta_{sg}$. The second model is a reduced model where $x$ is removed from the full model. The $p$-value for each gene set is obtained using the \emph{anova} function in R. For GSEA, we use the publicly available R package ({http://www.broad.mit.edu/gsea}) called R-GSEA. Maxmean, Mean.z and Mean.abs.z statistics are obtained using the GSA software available from \url{http://www-stat.stanford.edu/~tibs/GSA/index.html}. The computer programs for all the above models along with the R script to simulate data are available online at \url{http://ics.uci.edu/~babak/Codes}. 

In all six simulations, there are 1000 genes and 10 subjects (5 subjects are the control group and 5 subjects are the treatment group). The genes are randomly assigned to 50 gene sets. The number of genes in each set is itself random. The objective of the first simulation is to evaluate models based on data that conform with the normality assumptions for gene expression and random effects. To this end, we generate gene expression values by randomly sampling from $N(0, 1)$. For the first 5 gene sets, we randomly select a subset of genes to be differentially expressed. For each gene in this subset, we shift the expression values for the treatment group by a constant, which is randomly sampled from $N(0, 1)$. Note that the amount of shifts is different between genes, it could be positive (i.e., upregulated) or negative (i.e., downregulated), and it could be large or small. This procedure is repeated 100 times, each time a new dataset is generated. The above discussed methods are used for analyzing these datasets. We expect that each method ranks the first 5 sets as the most significant gene sets.

We follow the same procedure in Simulation 2, but this time, instead of sampling the expression values from $N(0, 1)$, we sample them from Gamma($a, 1$), where $a$ itself is sampled separately for each gene from the Uniform($1, 3$) distribution. This way, gene expressions are always positive with non-normal distributions that is skewed to the right. Moreover, the distributions of gene expression are different between genes. As before, the first 5 gene sets are designed to be differentially expressed; that is, we sample a subset of genes in these sets and add a random number from $N(0, 1)$ to the expression values for the treatment group.

For Simulation 3, we follow the same procedure as Simulation 2, but instead of sampling the amount change in expression for significant genes from $N(0, 1)$, we sample it from $\frac{1}{2}N(0, 0.5^{2})+\frac{1}{2}N(0, 1)$, which is a mixture of two normals. This way, the simulated data will not conform with the normality assumption for random effects in the mixed-effects model, and the normality assumption for the prior of $\beta$'s in our BGSA model. 

The objective of Simulation 4 is to evaluate the robustness of models when non-significant gene sets randomly include a highly significant gene. To this end, we start with the same procedure as Simulation 3. Then, we randomly select one third of non-significant gene sets (i.e., Sets 6-50), and shift the expression values for the treatment group by 2 for one of the genes. A robust model should not consider these gene sets as significant just because one of the genes is differentially expressed.   

Thus far, we assumed that the genes are independent. In reality, a gene could be correlated (or anti-correlated) with other genes. The objective of Simulation 5 is to examine the performance of models when the independence assumption among genes is violated. To this end, we follow the steps of Simulation 2, but this time, we randomly select some pairs of genes and make them correlated. The sampling is performed with replacement so it is possible that more than two genes become correlated with each other. The correlation coefficient itself is sampled randomly from Uniform(-1, 1). 

Finally, Simulation 6 is designed to evaluate the performance of the models when the sets share common genes. This scenario occurs in real studies quite often. For this simulation, we follow the same steps as Simulation 5 (where some genes are correlated with each other), but this time, we repeatedly select some of the genes (the number of genes selected is itself random) and add them to two or more gene sets. To make sure the first 5 gene sets remain relatively more significant compared to other gene sets, we select non-significant genes with a higher probability compared to the significant ones (0.95 vs. 0.05). It is still possible that some of the significant genes appear in the sets that are not significant overall. This would create a situation similar to that of Simulation 4.

For our BGSA model, we ran 4000 Markov chain Monte Carlo (MCMC) iterations, and obtained the posterior samples after discarding the first 500 iterations (i.e., pre-convergence iterations). The convergence of the model was evaluated based on the trace plots of hyperparameters ($\nu, \phi^{2}_{0}, \phi^{2}_{1}$ and $\tau^{2}_{s}$). 

The performance of each model is evaluated using the ROC curve which allows for simultaneous consideration of power (i.e., sensitivity) and type I error (i.e., $1$-specificity) without setting an arbitrary cut-off for significance level. Each point on the curve represents the number of correctly identified significant gene sets (vertical axis) compared to the number of times a model identified a non significant gene set as significant (horizontal axis) for different cut-offs. A more accurate model will have an ROC curve further away from the diagonal line (random model) with perfect prediction corresponding to the (0, 1) point. The Area under the ROC curve (AUC) is used as a summary statistic to compare models. For a perfect model, the AUC is equal to 100\%. For each simulation, we regard Sets 1-5 as true positive and all other sets as true negative. 

Table \ref{simRes} shows the average (over 100 repetitions) of AUC values for alternative methods under different simulation settings. The corresponding standard errors are presented in parentheses. For Simulation 1, LME outperforms all other models. This was expected since the data in this simulation are generated exactly according to the assumptions of LME (i.e., gene expressions are normal, the random effects are normal, and $\tau^{2}_{s}=0$ under the null). The BGSA model has the second best performance. Both LME and BGSA are significantly better than Maxmean, GSEA, Mean.z and Mean.abs.z (using paired $t$-tests, all the $p$-values are less than $0.001$). In the rest of simulations, the BGSA model provides the best performance (the improvements over all other models are statistically significant with $p$-values less than 0.01). The violation of model assumptions (especially when the gene sets share common genes) has a more sever impact on LME compared to BGSA. Finally, our results shows that Maxmean outperforms GSEA in all six simulations. This is consistent with the findings of \cite{efron07}.

\begin{table*}
 \centering
 \def\~{\hphantom{0}}

 \caption{Comparing our hierarchical Bayesian models to alternative approaches based on the average (presented in percentage) of area under the ROC curve (AUC) over 100 data sets for each simulation. The numbers in parentheses show the corresponding standard error for the average AUC.}
 \vspace*{6pt}
 
\label{simRes} 
\begin{center}
\begin{tabular*}{\columnwidth}{l | c c c c c c }
 & GSEA &  Mean.z & Mean.abs.z & Maxmean & LME & BGSA \\
\cline{1-7}
Simulation 1 & 66.1 (0.95)  &  64.9 (0.94) &   74.7 (1.26) &  81.4 (1.17) &   {\bf{89.1}}  (0.94) &  86.5 (1.08)   \\
Simulation 2& 60.4 (0.77) &   65.0 (0.98) &   69.6 (1.15)  &  77.2 (1.04) &  77.8 (1.06) &    {\bf{82.5}}  (1.02) \\
Simulation 3& 61.8  (0.77) &  61.1  (0.82) &  66.0  (1.01) &  68.8 (1.11) &   71.0  (1.14) &   {\bf{74.1}} (1.19)  \\
Simulation 4& 61.2 (0.78) &   62.6 (0.89) &  60.9  (0.81)  & 65.0 (1.04) &  65.1 (1.02) & {\bf{67.5}} (0.92)\\
Simulation 5& 61.8 (0.89) &   63.2 (0.93) &   70.1 (1.17)  &  76.0 (1.19) &  78.0 (1.22) &    {\bf{83.1}}  (1.15)\\
Simulation 6& 62.7 (0.81) &   65.6 (0.99) &   67.6 (1.02)  &  76.8 (1.18) &  74.9 (1.17) &    {\bf{81.4}}  (1.21)                                    
\end{tabular*}
\end{center}
\vspace*{-6pt}
\end{table*}

\section{Results for experimentally-derived data}
We apply our method to an experimentally-derived data set obtained based on interrogating the mutation status of p53 in cancer cell lines from the NCI-60 collection \citep{ross00}. This collection was created to explore gene expression in 60 cell lines using DNA microarrays prepared by robotically spotting 9,703 human cDNAs on glass microscope slides. The cDNAs included approximately 8,000 different genes. \cite{subramanian05} used these data to identify targets of the transcription factor p53, which regulates gene expression in response to various signals of cellular stress. They found the mutational status of the p53 gene for 50 of the NCI-60 cell lines using the IARC TP53 database \citep{oliver02}. Out of 50 cell lines ($n=50$), 17 were classified as normal and the remaining 33 were classified as carrying mutations in the gene. 
 
Using the curated gene sets available from the Molecular Signatures Database (MSigDB), we allocate genes into 522 gene sets. We use our BGSA to identify the gene sets that are significantly associated with p53 mutation status. The MCMC simulation ran for 2000 iterations of which the first 500 were discarded. Table \ref{pathways} shows the selected pathways by setting the cutoff for $P(v_{s}=0|D)$ at 0.1. While most of these gene sets are also identified as significant by Maxmean and GSEA, there are notable differences between the results based on these methods. For example, Cell Cycle Regulator, ATM, and BAD pathways, which appear in Table \ref{pathways}, are not selected as significant by \cite{efron07} and \cite{subramanian05}. On the other hand, some pathways such as RAS and NGF are identified as significant by GSA but not by BGSA. 

Both the ataxia telangiectasia gene (ATM) and BAD are strongly involved with p53. ATM encodes a protein kinase that acts as a tumor suppressor. ATM activation, via IR damage to DNA, stimulates DNA repair and blocks cell cycle progression. One mechanism through which this occurs is ATM dependent phosphorylation of p53, which causes growth arrest of the cell at a checkpoint to allow for DNA damage repair or can cause the cell to undergo apoptosis if the damage cannot be repaired. BAD, as a member of the BCL2 family, is also involved with the p53 pathway. BAD physically interacts with cytoplasmic p53 thereby preventing p53 from entering the nucleus. BAD mRNA and protein are increased in response to the upregulated expression of wild-type p53 but not the mutant p53 in cell lines. 

\begin{table*}
 \centering
 \def\~{\hphantom{0}}

\caption{List of significant gene sets in p53 dataset based on our BGSA model. The sets are selected by setting the cutoff for $P(v=0|D)$ at 0.1.}

 \vspace*{6pt}
 
\begin{minipage}{125mm}

\label{pathways}

\begin{tabular*}{\columnwidth}{l  c  c  c}
Gene set & Number of genes  & $E(\tau^2|D)$ & $P(v=0|D)$\\
\cline{1-4}
p53  Pathway & $16$   & 0.34 & 0.004 \\
Radiation Sensitivity & $26$  & 0.26 & 0.004\\
p53 Hypoxia Pathway & $20$  & 0.28 & 0.011 \\
p53 Up & $40$   & 0.28 & 0.016\\
Cell Cycle Regulator & $23$ & 0.19 & 0.023 \\
DNA Damage Signaling & 90 & 0.15 & 0.047 \\
ATM Pathway & 19 & 0.19 & 0.079 \\
g2 Pathway & 23 & 0.17 & 0.091\\
BAD Pathway & 21 & 0.15 & 0.098
\end{tabular*}

\end{minipage}
\vspace*{-6pt}
\end{table*}

One main concern about gene set analysis methods is how the size of gene sets influences the result. For our method, there is no strong relationship  (the correlation coefficient is 0.07) between the number of genes in a gene set and $P(v_s = 0 |D)$.

\section{Discussion and future directions}
We propose a new method for evaluating the significance of gene sets using a hierarchical Bayesian model. While simulation studies show that our approach outperforms other methods, such as GSEA and GSA, there are ways to further improve our model.

Compared to all other approaches discussed in this paper, our BGSA model is more computationally intensive. Using an implementation of R script on a Linux CentOS 5.3 machine with 2191 MHz processor speed, 2000 iterations of posterior sampling for p53 dataset takes approximately 20 minutes to run. This is much slower than, for example, the GSA model, which runs for 25 seconds on the same dataset (with 100 permutations). 

In this paper, we assumed that the samples are independent. Our model could be modified for grouped observations. To do this, we can simply add another random intercept, $\omega_{sgi}$, that varies from one observation group to the other; that is we modify our model as follows:
\begin{eqnarray*}\label{model1}
y_{sgij} & = & \alpha_{sg}+ \omega_{sgi} + \beta_{sg} x_{ij} + \epsilon_{sgij}, \qquad g=1, 2, \ldots, \ell_{s},
\end{eqnarray*}
where $i$ is now the index for to the $i^{th}$ group of observations, and $j$ is the index for observations within the group. The posterior sampling and inference remain as before.

Our hierarchical Bayesian model could also be extended for problems with multiple samples (e.g., experimental conditions). For example, the treatment group might be divided into different treatment options. For such problems, we can modify the prior for $\beta$ as follows:
\begin{eqnarray*}
\beta_{sgc} & \sim & N(0, \tau^{2}_{s}), \qquad c=1, \ldots, C,
\end{eqnarray*}
where $c$ is the index for the class group (e.g., experimental conditions), and $\beta_{sgc}$ is the random effect parameter associated with class $c$. Note that the control group with $c=0$ is considered as the baseline, and $x_i$ is now a vector of dummy variables. In this setting, all $\beta$'s in a set are still controlled by one hyperparameter, $\tau^{2}_{s}$. Alternatively, we could have an additional hyperparameter for each gene
\begin{eqnarray*}
\beta_{sgc} & \sim & N(0, \tau^{2}_{s}\rho^{2}_{sg})
\end{eqnarray*}
This way, $\rho_{sg}$ controls all $\beta$'s associated with gene $g$ in set $s$. This setting makes it easier for the variance of $\beta_{sg}$ to shrink towards zero if a gene is not significant (i.e., its corresponding $\beta$'s are small) while other genes in the set seem to be significant. If the set as a whole is not signifiant, the posterior distribution of $\tau_{s}$ will shrink towards zero to make all $\beta$'s (that are associated with set $s$) small simultaneously.

Currently, our method does not take into account possible overlaps between gene sets. In reality, the gene sets are not mutually exclusive. Our approach could easily be extended to take into account the overlaps between gene sets. We could, for example, modify our prior such that gene sets that include common genes share common hyperparameters. Without loss of generality, consider two gene sets $\mathcal{S}_{s}=\{G_{1}, G_{2}, \ldots, G_{q}, G_{s, q+1}, G_{s, q+2}, \ldots, G_{s\ell_{s}}\}$ and $\mathcal{S}_{r}=\{G_{1}, G_{2}, \ldots, G_{q}, G_{r, q+1}, G_{r, q+2}, \ldots, G_{r\ell_{r}}\}$ that share $q$ genes: $\{G_{1}, \ldots, G_{q}\}$. We can use the following priors for $\beta$'s to account for the overlap:
\begin{eqnarray*}
\beta_{1}, \ldots, \beta_{q} | \tau_{rs} & \sim & N(0, \tau^{2}_{rs})\\
\beta_{s, q+1}, \ldots, \beta_{s\ell_{s}} | \tau_{rs}, \tau_{s} & \sim & N(0, \tau^{2}_{s} + \tau^{2}_{rs})\\
\beta_{r, q+1}, \ldots, \beta_{r\ell_{r}} | \tau_{rs}, \tau_{r} & \sim & N(0, \tau^{2}_{r} + \tau^{2}_{rs})
\end{eqnarray*}
This way, if the common genes are not significant, their corresponding hyperparameter, $\tau^{2}_{rs}$, shrinks towards zero, and the significance of sets $\mathcal{S}_{s}$ and $\mathcal{S}_{r}$ would depend solely on the non-overlapping genes. However, if the common genes are in fact significant, $\tau^{2}_{rs}$ moves away from zero and contributes to the overall significance of both gene sets. 

The method proposed in this paper regards the gene sets as known and fixed, as is done in GSA and GSEA. Therefore, none of these methods take into account the uncertainty regarding the grouping of genes. In reality, there are many ways to group genes depending on which biological aspect we consider. A similar concern is recently discussed by \cite{shen08}. Our hierarchical Bayesian model could address this issue by assuming multiple grouping schemes each with its own separate set of hyperparameters. 

While the main motivating application for our proposed model is identifying signaling pathways (each pathway is associated with a collection of interconnected genes), our method could be applied to a wide range of problems, for which a large number of hypotheses are grouped into subsets of related hypotheses according to some prior information. One such problem is functional neuroimaging for human brain mapping, to study normal versus pathological brain functions. A typical experiment in this area involves assessing a large number of pixels (each pixel representing a small area of brain tissue). It is possible to group pixels such that each subset of pixels represents a different brain region. Another possible application for our method is analysis of single-nucleotide polymorphisms (SNPs) in genome-wide association studies, in which the objective is to identify and characterize genetic variants related to common complex diseases. These studies commonly focus on haplotypes (a set of statistically associated SNPs that are transmitted together as a block) on a single chromatid. More recently, there has also been suggestions to focus on pathways in genome-wide association studies \citep[see for example, ][]{luan08}.

\section*{Acknowledgements}
The authors also gratefully acknowledge the Young Clinical Scientist Award from FAMRI and NIH funding from U54CA11967 and U56CA112973. 

 \bibliographystyle{../Chicago}

\end{document}